# Network Coding Security: Attacks and Countermeasures

Luísa Lima, João P. Vilela, Paulo F. Oliveira and João Barros

Instituto de Telecomunicações
Departamento de Ciência de Computadores
Faculdade de Ciências da Universidade do Porto
Rua do Campo Alegre, 1021/1055, 4169-007 Porto, Portugal
Email: {luisalima, joaovilela, pvf, barros}@dcc.fc.up.pt

## ABSTRACT

By allowing intermediate nodes to perform non-trivial operations on packets, such as mixing data from multiple streams, network coding breaks with the ruling store and forward networking paradigm and opens a myriad of challenging security questions. Following a brief overview of emerging network coding protocols, we provide a taxonomy of their security vulnerabilities, which highlights the differences between attack scenarios in which network coding is particularly vulnerable and other relevant cases in which the intrinsic properties of network coding allow for stronger and more efficient security solutions than classical routing. Furthermore, we give practical examples where network coding can be combined with classical cryptography both for secure communication and secret key distribution. Throughout the paper we identify a number of research challenges deemed relevant towards the applicability of secure network coding in practical networks.

## Categories and Subject Descriptors

C.2.1 [**Computer-Communication Networks**]: Network Architecture and Design—*Network Communications*; C.2.2 [**Computer-Communication Networks**]: Network Protocols—*Applications*; E.3 [**Data**]: Data Encryption—*Code Breaking*; E.4 [**Data**]: Coding and Information Theory—*Error Control Codes*

## General Terms

Fundamentals and Research Challenges

## 1. INTRODUCTION

Disruptive technologies are often at the heart of novel security concerns and solutions. In the context of communication networks and protocols, network coding [Ahlswede et al. 2000] offers an arguably intriguing networking concept: data throughput and network robustness can be considerably improved by allowing the intermediate nodes in a network to mix different data flows through algebraic combinations of multiple datagrams.

This key idea, which clearly breaks with the standard store-and-forward paradigm of current routing solutions, is illustrated in *Figure 1*. To exchange messages $a$ and $b$, nodes $A$ and $B$ must route their packets through node $S$. Clearly, the traditional scheme shown on top would require four transmissions. However, if $S$ is allowed to perform network coding with simple XOR operations, as illustrated in the lower diagram, $a \oplus b$ can be sent in one single broadcast transmission (instead of one transmission with $b$ followed by another one with $a$). By combining the received data with the stored message, $A$ which possesses $a$ can recover $b$ and $B$ can recover $a$ using $b$. Thus, in this scenario, network coding saves one transmission. More sophisticated network coding protocols view packets as a collection of symbols from a particular finite field and forward linear combinations of these symbols across the network, thus leveraging basic features of linear codes such as erasure correction capability and well understood encoding and decoding algorithms.

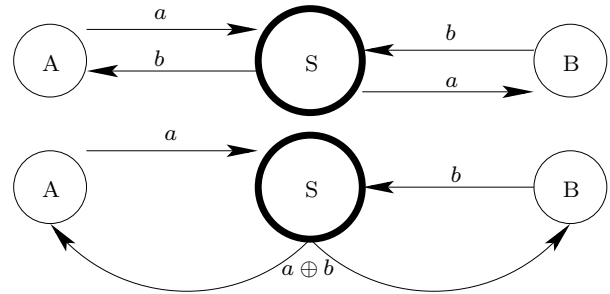

**Figure 1: A typical example of wireless network coding.**

The main goal of this survey paper is to present and discuss the most salient aspects of network coding from the point of view of network security. First, we provide a taxonomy of network coding vulnerabilities and attacks, and compare them against those of state-of-the-art routing algorithms. Some emphasis shall be given to active attacks, which can lead to severe degradation of network coded information flows. Then, we show how to leverage the intrinsic properties of network coding for information security and secret key distribution, in particular how to exploit the fact that nodes observe algebraic combinations of packets instead of the data packets themselves. Although the prevalent design methodology for network protocols views security as something of an add-on to be included after the main communications tasks have been addressed, we shall contend that the special characteristics of network coding warrant a more comprehensive approach, namely one which gives equal importance to security concerns. In the process, we elaborate on a number of open problems for future research.

The remainder of the paper is organized as follows. Section 2 provides an overview of currently available practical network coding schemes and proposes a simple classification for security purposes. The main security challenges of network coding are explained in Section 3, followed by an overview of specific solutions, which are presented in Section 4. The paper is concluded in Section 5.

## 2. PRACTICAL NETWORK CODING PROTOCOLS

Although network coding is a fairly recent technology, there exist already a number of protocol proposals which explore network coding for higher throughput or robustness in various applications and communication networks. From the point of view of network security, it is useful to divide network coding protocols into two main classes:

1. *stateless network coding protocols*, which do not rely on any form of state information to decide when and how to mix different packets in the sender queue;

2. *state-aware network coding protocols*, which rely on partial or full network state information (e.g. buffer states of neighboring nodes, network topology, or link costs) to compute a network code or determine opportunities to perform network coding in a dynamic fashion.

As shall be demonstrated in Section 3, the security vulnerabilities of the protocols in the first and the second class are quite different from each other, most notably because the former require state information and node identification to be disseminated in the network and are thus more vulnerable to a wide range of impersonation and control traffic attacks.

### 2.1 Stateless Network Coding Protocols

A representative for the first class, Random Linear Network Coding (RLNC) is a completely distributed methodology for combining different information flows [Ho et al. 2004]. The basic principle is that each node in the network selects a set of coefficients independently and randomly and then sends linear combinations of the data symbols (or packets) it receives. *Figure 2* illustrates the linear operations carried out at intermediate node $v$ (using integers for simplicity). The symbols $x$, $y$ and $z$ denote the native packets, which convey the information to be obtained at the receivers via Gaussian elimination. $P1$ and $P2$ arrive at intermediate node $v$ in the network through its incoming links. $P3$, which is sent through the only outgoing link of node $v$, is the result of a random linear combination of $P1$ and $P2$ at node $v$, with chosen coefficients 1 and 2, respectively.

The *global encoding vector*, i.e. the matrix of coefficients which holds the linear transformations that the original packet goes through on its path from the source to the destination, is sent along in the packet header to ensure that the end receivers are capable of decoding the original data. Specifically, it was shown that if the coefficients are chosen at random from a large enough field, then Gaussian elimination succeeds with overwhelming probability [Ho et al. 2003, Ho et al. 2004]. It has also been shown that RLNC is capacity-achieving, even on asynchronous lossy packet networks [Lun et al. 2005].

A framework for packetized network coding (*Practical Network Coding*, PNC) is presented in [Chou et al. 2003], which leverages RLNC's resilience against disruptions such as packet loss, congestion, and changes of topology, in order to guarantee robust communication over highly dynamic networks with minimal (or no) control information. The framework defines a packet format and a buffering model. The packet format includes the *global encoding vector* in its header. The payload of the packets is divided into vectors according to the field size ($2^8$ or $2^{16}$, i.e. each symbol has 8 or 16 bits, respectively). Each of these symbols is then used as a building block for the linear operations performed by the nodes.

The buffering model divides the stream of packets into *generations* of size $h$, such that packets in the same generation are tagged with a common generation number. Each node sorts the incoming packets in a single buffer according to their generation number. When there is a transmission opportunity at an outgoing edge, the sending node generates a new packet, which contains a random linear combination of all packets in the buffer that belong to the *current* generation. If a packet is *non-innovative*, i.e. if it does not increase the rank of the decoding matrix available at the receiving node, then it is immediately discarded. As soon as the matrix of received packets has full rank, Gaussian elimination is performed at the receivers to recover the original packets.

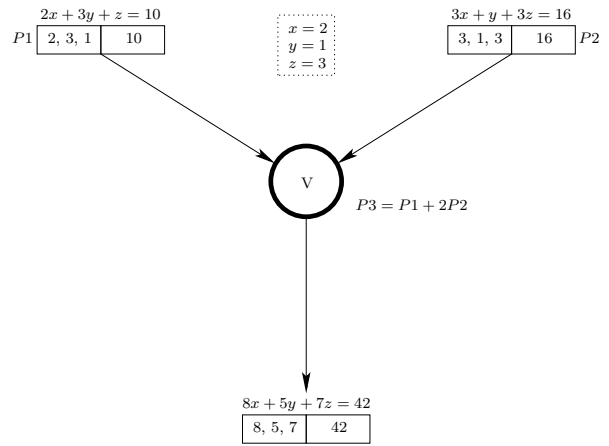

**Figure 2: An instance of random linear network coding.**

RLNC seems particularly beneficial in dynamic and unstable networks – that is, networks where the structure or topology of the network varies within a short time, such as mobile ad-hoc networks and peer-to-peer content distribution networks. The benefits of RLNC in wireless environments with rare and limited connectivity, either due to mobility or battery scarcity, are highlighted in [Widmer and Boudec 2005], which proposes an algorithm aimed at reducing the overhead of probabilistic routing algorithms with applications in delay tolerant networks. Other RLNC based information dissemination schemes that increase reliability and robustness while reducing the incurred overhead can be found in [El Fawal et al. 2006, Deb et al. 2005].

The potential impact of RLNC in content distribution networks has been addressed in [Gkantsidis and Rodriguez 2005, Dimakis et al. 2007]. Since each node forwards a random linear combination independently of the information present at other nodes, its operation is completely decentralized. Moreover, when collecting a random combination of packets from a randomly chosen node, there is high probability of obtaining a linearly independent packet in each time. Thus, the problem of redundant transmissions, which is typical of traditional flooding approaches, is considerably reduced. In content distribution applications, there is no need to download one particular fragment; instead, any linearly independent segment brings innovative information [Dimakis et al. 2007]. The practical scheme proposed in [Gkantsidis and Rodriguez 2005] is designed for large files and nodes make forwarding decisions solely based on local information. It is shown that network coding not only improves the expected file download time, but also improves the overall robustness of the system with respect to frequent changes in the network.

Acedanski *et al* [Acedanski et al. 2005] compare the performance of network coding with traditional coding measures in a distributed storage setting in which each storage location only has very limited storage space for each file and the objective is that a file-downloader connects to as few storage locations as possible to re-trieve a file. It is shown that RLNC performs well without the need for a large amount of additional storage space required at a centralized server. Dimakis *et al* [Dimakis et al. 2007] investigate how to engineer distributed storage systems from a coding perspective, in particular for peer-to-peer scenarios. Their contribution includes a general graph-theoretic framework for computing lower bounds on the bandwidth required to maintain distributed storage architectures and shows that RLNC achieves these lower bounds.

## 2.2 State-aware Network Coding Protocols

State-aware network coding protocols rely on the available state information to optimize the coding operations carried out by each node. The optimization process may target the throughput or the delay (among other performance metrics). Its scope can be *local* or *global*, depending on whether the optimization affects only the operations within the close neighborhood of a node or addresses the end-to-end communication across the entire network. For instance, if control traffic is exchanged between neighbors, each node can perform a local optimization to decide on the fly how to mix and transmit the received data packets. One way to implement protocols with global scope is to use the polynomial time algorithm of [Jaggi et al. 2005], which, given the network graph, is able to determine the optimal network code prior to communication. By exploiting the broadcast nature of the wireless medium and spreading encoded information in a controlled manner, state-aware protocols promise considerable advantages in terms of throughput, as well as resilience to node failures and packet losses. Efficiency gains come mainly from the fact that nodes make use of every data packet they overhear and, in some instances, also from a reduced need for control information.

As an example, the COPE protocol [Katti et al. 2006] inserts a coding layer between the IP and MAC layers for detecting coding opportunities. More specifically, nodes overhear and store packets that are exchanged within their radio range, and then send reception reports to inform their neighbors about which packets they have stored in their buffers. Based on these updates, each node computes the optimal XOR mixture of multiple packets in order to reduce the number of transmissions. It has been shown that this approach can lead to strong improvements in terms of throughput and robustness to network dynamics. The interplay of routing and network coding is explored in [Toledo and Wang 2006, Sengupta et al. 2007].

## 3. SECURITY CHALLENGES

The aforementioned network coding protocols expect a well behaved node to obey the following rules:

- *encode the received packets correctly*, thus contributing to the expected benefits of network coding;

- *forward the encoded packets correctly*, thus enabling the destination nodes to retrieve the intended information;

- *ignore data for which it is not the intended destination*, thus fulfilling basic confidentiality requirements.

In the case of state-aware network coding protocols, we must add one more rule:

- *participate in the timely dissemination of correct state information*, thus contributing to a sound knowledge base for network coding decisions.

It follows that an attack on a network coding based protocol must result from one or more network nodes breaking one or more of these basic rules.

In *Table 1* we provide a taxonomy of security vulnerabilities and their impact in both the traditional network paradigm and network coding based protocols. Our goal is not to provide an exhaustive list but rather to emphasize those that underline the effectiveness of network coding based protocols in protecting against some typical attacks on communication networks, and also those that may cause considerable damage particularly to this type of protocols. Naturally, the means to achieve a successful attack are of course highly dependent on the specific rules of the protocol, and therefore it is reasonable to distinguish between the aforementioned stateless and state-aware classes of network coding schemes.

If properly applied, stateless network protocols based on RNLC are potentially less subject to some of the typical security issues of traditional routing protocols. First, stateless protocols do not depend on exchange of topology or buffer state information, which can be faked (e.g. through *link spoofing* attacks). Secondly, the impact of *traffic relay refusal* is reduced, due to the inherent robustness that results from spreading the information by means of network coding. Thirdly, the information retrieval depends solely on the data received and not on the identity of nodes, which ensures some protection against *impersonation* attacks.

In contrast, state-aware network coding protocols rely on vulnerable control information disseminated among nodes in order to optimize the encodings. On one hand, this property renders them particularly prone to attacks based on the generation of false control information. On the other hand, control traffic information can also be used effectively against active attacks, such as the injection of erroneous packets. For protocols with local scope, the negative impact of active attacks is limited to a confined neighborhood, whereas with end-to-end network coding the consequences can be much more devastating. Opportunistic network coding protocols, which rely on the information overheard by neighboring nodes over the wireless medium, are obviously more amenable to eavesdropping attacks than their wireline counterparts.

More generally, in comparison with traditional routing, the damage caused by a malicious (Byzantine) node injecting corrupted packets into the information flow is likely to be higher with network coding, irrespective of whether a protocol is stateless or state-aware. Since network coding relies on mixing the content of multiple data packets, a single corrupted packet may very easily corrupt the entire information flow from the sender to the destination at any given time.

## 4. SECURING NETWORK CODING PROTOCOLS

Having provided a security taxonomy evidencing the specific vulnerabilities of network coding, we now turn our attention to the next natural question, namely how to find appropriate mechanisms for securing network coding protocols. Our main goal here is to show how the specific characteristics of network coding can be leveraged to counter some of the threats posed by eavesdroppers and Byzantine attackers. To further illustrate its security potential, we include a mobile key distribution scheme in which network coding adds an extra line of defense.

## 4.1 Countering Eavesdropping Attacks

We shall start by presenting countermeasures against passive attacks, with special emphasis on three different scenarios. First, we shall consider nice but curious nodes, which do not break any of the established rules except for not ignoring the data for which they are not the intended receivers. In the second instance, the eavesdropper is able to wiretap a subset of network links. Finally, the third type of attacker is a worst-case eavesdropper who is given full access to all the traffic in the network.

**Table 1:** Security vulnerabilities of Stateless and State-Aware Network Coding (NC) Protocols in Comparison with Traditional Routing

| ATTACK | DESCRIPTION | TRADITIONAL ROUTING | STATELESS NC | STATE-AWARE NC |
|---|---|---|---|---|
| Impersonation | A node generates messages pretending to be another node. | By generating routing messages pretending to be another node, the attacker can introduce conflicting routes or routing loops, and cause network partitioning. | Stateless NC protocols do not rely on the identity of the nodes for any operation, therefore they are not affected. | State-aware NC protocols rely on network nodes for state information and, thus, identities can be faked to convey wrong information. |
| Byzantine Fabrication | A node generates messages containing false information. | By generating routing messages with false information (e.g. announcing small distances to nodes that are far away), the attacker can cause degradation of communications and traffic interception. | Stateless NC protocols can be affected in terms of the gains obtained and the processing time by the injection of erroneous packets into the information flow. | State-aware NC protocols are also prone to this type of attack. If, on one hand, one extra type of information (control traffic) can be targeted, on the other hand, that same information may be used to provide a faster detection of such attacks and prevent dissemination of bogus packets. |
| Byzantine Modification | A node modifies the messages in transit. | By changing the header fields of messages passed among nodes (e.g. the destination node), the attacker can cause traffic subversion and denial of service. | Stateless NC protocols can be affected by changes in the coded packets in transit, in particular by changes in the coefficients and/or the encoded payload which may render the native packets undecodable. | State-aware NC protocols can be affected by changes on either the coded data packets or the control packets. Also here the control packets can possibly help in addressing these attacks. |
| Byzantine Replay attack | A node sends "old" previously transmitted (and eventually authenticated) messages to the network. | By sending "old" routing messages, outdated, conflicting and/or wrong information enters the network which may cause defective routing. | Stateless NC protocols can be affected in terms of NC gain and processing time by the injection of erroneous packets which are repeated into the information flow. | State-aware NC protocols can be affected by changes on either the coded data packets or the control packets. Also here the control packets can possibly help in addressing these attacks. |
| Blackhole | A node refuses to relay traffic on behalf of others. | The attacker can cause denial of service and degradation of communications. | Stateless NC protocols can be affected by degradation of communications however they benefit from the inherent redundancy of NC enabling increased robustness and improved probability of successful delivery [Ho et al. 2004]. | State-aware NC protocols can be affected by degradation of communications, but can possibly benefit from the redundancy of this type of protocols. |
| Eavesdropping (internal or external) | The nodes collaborate with the protocol, however they try to acquire as much information as possible. | By looking at every message that a node is expected to relay, the attacker can get access to classified information. | If an intermediate node has access to a sufficient number of linearly independent combinations of packets, it can decode them and have access to all of the sent information. | State-aware NC protocols are also prone to eavesdropping threats. Control traffic may be used both to prevent as well as potentiate this type of attack. |

### 4.1.1 Nice But Curious Nodes

Consider first a threat model in which the network consists entirely of *nice but curious* nodes, i.e. they comply with the communication protocols (in that sense, they are well-behaved) but may try to acquire as much information as possible from the data flows that pass through them (in which case, they are potentially ill intended). Under this scenario, stateless protocols that exploit the RLNC scheme described in Section 2 possess an intrinsic security feature [Lima et al. 2007]: depending on the size of the code alphabet and the topology of the network it is in many instances unlikely that an intermediate node will have enough degrees of freedom to perform Gaussian elimination and gain access to the transmitted data set.

Based on this observation, it is possible to characterize the threat level posed by an intermediate node according to an *algebraic security criterion* [Lima et al. 2007] that takes into account the number of components of the global encoding vector it receives. In the example of *Figure 3*, which uses integers for simplicity, the upper (uncoded) transmission scheme leaves partial data unprotected, whereas in the lower (network coding) scheme the intermediate nodes 2 and 3 are not able to recover the data symbols.

### 4.1.2 Wiretapping Nodes

A different threat model, commonly found in the recent literature on secure network coding, assumes that one or more external eavesdroppers (or wiretappers) have access to a subset of the available communication links. The crux of the problem is then to find code constructions capable of splitting the data among different links in such a way that reconstruction by the attackers is either very difficult or impossible. Under this assumption, it was shown in [Cai and Yeung 2002], that there exist secure linear network codes that achieve perfect information-theoretic secrecy for single source multicast. In [Cai and Yeung 2007], these results are generalized to multi-source linear network codes by using the algebraic structure of such codes to derive necessary and sufficient conditions for their security. Furthermore, [Yeung and Cai 2008] goes on to prove the optimality of a secure network code proposed in [Cai and Yeung 2002]. More specifically, it is shown that the constructed code maximizes the amount of secure multicasted information while minimizing the necessary amount of randomness. The contribution of [Silva and Kschischang 2008] consists of a coding scheme that can achieve the maximum possible rate of $n-\mu$ information theoretically secure packets, where $n$ is the number of packets from source to each receiver and $\mu$ is the number of links that the wiretapper can observe. This can be applied on top of any communication network without requiring any knowledge of the underlying network code and without imposing any coding constraints. The basic idea is to use a "nonlinear" outer code, which is linear over an extension field $\mathbb{F}_{q^m}$, and to exploit the benefits of this extension field. Some contributions propose a different criterion in which a system is deemed to be secure if an eavesdropper is unable to get any uncoded or immediately decodable (also called *meaningful*) source data [Bhattad and Narayanan 2005]. Other contributions exploit the network topology to ensure that an attacker is unable to get any meaningful information and add a cost function to the secure network coding problem. The problem then becomes finding a coding scheme that minimizes both the network cost and the probability that the attacker is able to retrieve all the messages of interest [Tan and Médard 2006].

Since state-aware network coding protocols with local code optimization (e.g. COPE [Katti et al. 2006]) expect neighboring nodes to be able to decode all the packets they receive, confidentiality must be ensured by means of end-to-end encryption.

### 4.1.3 Worst-Case Eavesdroppers

In this case, the threat model is one in which the attacker has access to all the packets traversing the network but not to the secret keys shared among legitimate communicating parties. SPOC (Secure Practical Network Coding) [Vilela et al. 2008] is a light-weight security scheme for confidentiality in RLNC which provides a simple yet powerful way to exploit the inherent security of RLNC in order to reduce the number of cryptographic operations required for confidential communication. This is achieved by protecting (or "locking") only the source coefficients required to decode the linearly encoded data, while allowing intermediate nodes to run their network coding operations by the means of "unlocked" coefficients which provably do not compromise the hidden data. The latter set of coefficients stores the operations performed along the network upon the packet.

Seeking to evaluate the level of security provided by SPOC, [Lima et al. 2008] analyzes the mutual information between the encoded data and the two components that can lead to information leakage, namely the matrices of random coefficients and the original data itself. This analysis, which is independent of any particular cypher used for locking the coefficients, assumes that the encoding matrices are based on variants of random linear network coding and can only be accessed by the source and sinks. The results, some of which hold even with finite block lengths, prove that information-theoretic security is achievable for any field size without loss in terms of decoding probability. In other words, since correlation attacks based on the encoded data become impossible, protecting the encoding matrix is generally sufficient to ensure the confidentiality of network coded data.

## 4.2 Countering Byzantine Attacks

Although Byzantine attacks can have a severe impact on the integrity of network coded information, the specific properties of linear network codes can be used effectively to counteract the impairments caused by traffic relay refusal or injection of erroneous packets. In particular, RLNC has been shown to be very robust to packet losses induced by node misbehavior [Lun et al. 2005]. More sophisticated countermeasures, which modify the format of coded packets, can be subdivided into two main categories: (1) end-to-end error correction, and (2) misbehavior detection, which can be carried out either packet by packet or in generation based fashion.

### 4.2.1 End-to-end error correction

The main advantage of *end-to-end error-correcting codes* is that the burden of applying error control techniques is left entirely to the source and the destinations, such that intermediate nodes are not required to change their mode of operation. The typical transmission model for end-to-end network coding is well described by a matrix channel $\mathbf{Y} = \mathbf{AX} + \mathbf{Z}$, where $\mathbf{X}$ corresponds to the matrix whose rows are the transmitted packets, $\mathbf{Y}$ is the matrix whose rows are the received packets, $\mathbf{Z}$ denotes the matrix corresponding to the injected error packets after propagation over the network, and $\mathbf{A}$ describes the transfer matrix, which corresponds to the global linear transformation operated on packets as they traverse the network. In terms of performance, error-correction schemes can correct up to the min-cut between the source and the destinations. Rank-metric error-correcting codes in RLNC under this setting appear in [Koetter and Kschischang 2007] and the results are extended in [Silva et al. 2007; Silva and Kschischang 2007] for the scenario in which the channel may supply partial information about erasures and deviations from the sent information flow. Still under the same setting, [Silva et al. 2008] considers a probabilistic error model for random network coding, provides bounds on capacity and presents a simple coding scheme with polynomial complexity that achieves capacity with exponentially low probability of failure with respect to both the packet length and the field size. [Montanari and Urbanke 2007] extends [Koetter and Ksch and provides an error-correction scheme based on random sparse graphs and a low-complexity decoding algorithm for a probabilistic error model. Bounds on the maximum achievable rate in an adversarial setting are provided by [Cai and Yeung 2003], which derives generalizations of the Hamming and Gilbert-Varshamov bounds.

A somewhat different approach to network error correction is considered in [Jaggi et al. 2007], which proposes robust network codes that have polynomial-time complexity and attain optimal rates in the presence of active attacks. The basic idea is to regard the packets injected by an adversarial node as a second source of information and adding enough redundancy to allow the destination to distinguish between relevant and erroneous packets. The achieved capacity depends on the rate at which the attacker can obtain information, as well as on the existence of a shared secret between the source and the sinks. This work is extended in [Nutman and Langberg 2008], which introduces three additional adversarial models and gives op-timal rates $C - z$ for each, where $C$ is the network capacity and $z$ is the number of links controlled by an attacker.

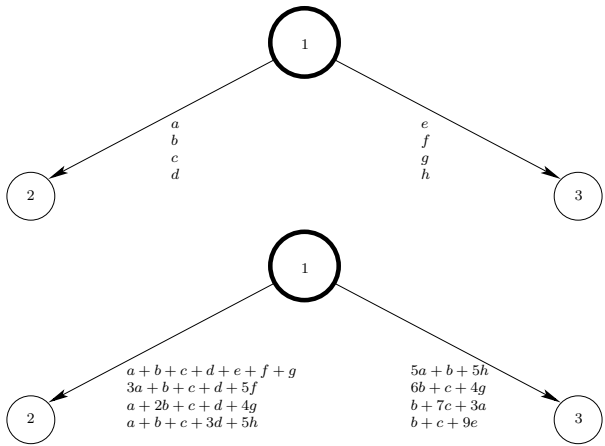

**Figure 3: Example of algebraic security. The top scheme discloses data to intermediate nodes, whereas the bottom scheme can be deemed algebraically secure.**

### 4.2.2 Misbehavior Detection

*Generation-based* detection schemes generally offer similar advantages as network error-correcting codes in that the often computationally expensive task of detecting the modifications introduced by Byzantine attackers is carried out by the destination nodes. The main disadvantage of generation-based detection schemes is that only nodes with enough packets from a generation are able to detect malicious modifications, and thus, usage of such detection schemes can result in large end-to-end delays. [Ho et al. 2004] proposes an information-theoretic framework for detecting Byzantine attackers. The underlying assumption is that the attacker cannot see the full rank of the packets in the network. It is shown that a hash scheme with polynomial complexity can be used without the need for secret key distribution. However, the use of a block code forces an a priori decision on the coding rate.

The key idea of *packet-based detection schemes* is that some of the intermediate nodes in the network can detect polluted data on the fly and drop the corresponding packets, thus retransmitting only valid data. However, packet-based detection schemes require active participation of intermediate nodes and are dependent on hash functions, which are generally computationally expensive. Alternatively, this type of attacks can be mitigated by signature schemes based on homomorphic hash functions. The use of homomorphic hash functions is specifically tailored for network coding schemes, since the hash of a coded packet can be easily derived from the hashes of previously encoded packets, thus enabling intermediate nodes to verify the validity of encoded packets prior to mixing them algebraically. Unfortunately, homomorphic hash functions are also computationally expensive.

[Charles et al. 2006] proposes a homomorphic signature scheme for network coding which is based on Weil pairing in elliptic curve cryptography. Homomorphic hash functions are also considered in the context of peer to peer content distribution with rateless erasure codes for multicast transfers in [Krohn et al. 2004]. With the goal of preventing both the waste of large amounts of bandwidth and the pollution of download caches of network clients, each file is compressed to a smaller hash value, with which receivers can check the integrity of downloaded blocks. Beyond its independence from the coding rate, the main advantage of this process is that it is less computationally expensive for large files than traditional forward error correction codes (such as Reed-Solomon codes).

One of the contributions of [Gkantsidis and Rodriguez 2006] is a cooperative security scheme for on-the-fly detection of malicious blocks injected in network coding based peer-to-peer networks. In order to reduce the cost of verifying information on-the-fly while efficiently preventing the propagation of malicious blocks, the authors propose a distributed mechanism where every node performs block checks with a certain probability and alerts its neighbors when a suspicious block is found. Techniques to prevent denial of service attacks due to the dissemination of alarms are also included in [Gkantsidis and Rodriguez 2006].

In [Zhao et al. 2007] the basic idea is to take advantage of the fact that in linear network coding any valid packet transmitted belongs to the subspace spanned by the original set of vectors. A signature scheme is thus used to check that a given packet belongs to the original subspace. Generating a signature that is not in the subspace yet passes the check is shown to be hard.

A comparison of the bandwidth overhead required by Byzantine error correction and detection schemes is provided in [Kim et al. 2008]. The intermediate nodes are divided into regular nodes and trusted nodes, and only the latter are given access to the public key of the Byzantine detection scheme in use. Under these assumptions, it is shown that packet-based detection is most competitive when the probability of attack is high, whereas a generation-based approach is the more bandwidth efficient when the probability of attack is low.

## 4.3 Key Distribution Schemes

The ability to distribute secret keys in a secure manner is an obvious fundamental requirement towards assuring cryptographic security. In the case of highly constrained mobile ad-hoc and sensor networks, key pre-distribution schemes [Eschenauer and Gligor 2002] emerge as a strong candidate, mainly because they require considerably less computation and communication resources than trusted party schemes [Perrig et al. 2002] or public-key infrastructures [Malan et al. 2004]. The main caveat is that secure connectivity can only be achieved in probabilistic terms, i.e. if each node is loaded with a sufficiently large number of keys drawn at random from a fixed pool, then with high probability it will share at least one key with each one of its neighboring nodes.

The contributions in [Oliveira et al. 2007] show that network coding can be an effective tool for establishing secure connections between low-complexity sensor nodes. In contrast with pure key pre-distribution schemes, it is assumed that a mobile node, e.g. a handheld device or a laptop computer, is available for activating the network and for helping to establish secure connections between nodes. By exploiting the benefits of network coding, it is possible to design a secret key distribution scheme that requires only a small number of pre-stored keys, yet ensures that shared-key connectivity is established with probability one and that the mobile node is provably oblivious to the distributed keys [Oliveira et al. 2007]. The basic idea of the protocol, which is illustrated in Fig. 4, can be summarized in the following tasks:

(a) prior to sensor node deployment:

 1. a large pool $P$ of $N$ keys and their $N$ identifiers are generated off-line;
 2. a different subset of $L$ keys drawn randomly from $P$ and the corresponding $L$ identifiers are loaded into the memory of each sensor node;
 3. a table is constructed with the $N$ key identifiers and $N$ sequences that result from performing an XOR of each key with a common protection sequence $X$;
 4. the table is stored in the memory of the mobile node;

(b) after sensor node deployment:

 1. the mobile node broadcasts HELLO messages that are received by any sensor node within wireless transmission range;
 2. each sensor node replies with a key identifier;
 3. based on the received key identifiers the mobile node locates the corresponding sequences protected by $X$ and combines them through an XOR network coding operation, thus canceling $X$ and obtaining the XOR of the corresponding keys;
 4. the mobile node broadcasts the resulting XOR sequence;
 5. by combining the received XOR sequence with its own key, each node can easily recover the key of its neighbor node, thus sharing a pair of keys which is kept secret from the mobile node.

Although the use of network coding hereby presented is limited to XOR operations, more powerful secret key distribution schemes are likely to result from using linear combinations of the stored keys.

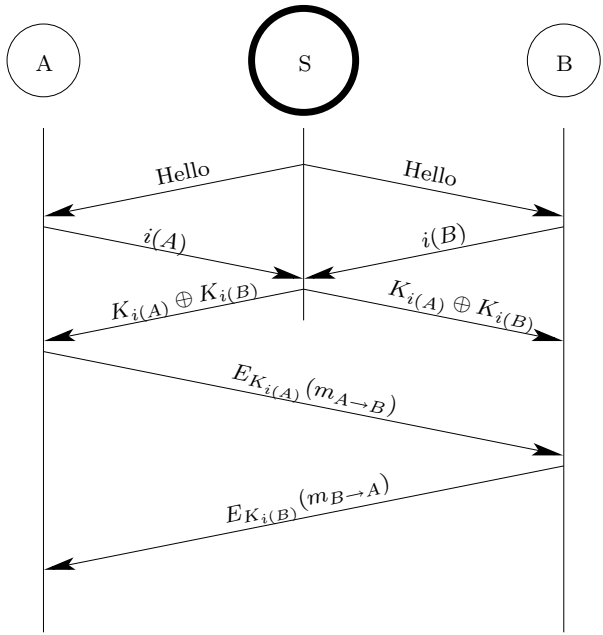

**Figure 4: Secret key distribution scheme. Sensor nodes $A$ and $B$ want to exchange two keys via an mobile node $S$. The process is initiated by an HELLO message broadcasted by $S$. Upon receiving this message, each sensor node sends back a key identifier $i(\cdot)$ corresponding to one of its keys $K_{i(\cdot)}$. Node $S$ then broadcasts the result of the XOR of the two keys $K_{i(A)} \oplus K_{i(B)}$. Once this process is concluded, sensor nodes $A$ and $B$ can communicate using the two keys $K_{i(A)}$ and $K_{i(B)}$ (one in each direction). Here, $E_{K_{i(A)}}(m_{A \to B})$ denotes a message sent by $A$ to $B$, encrypted with $K_{i(A)}$, and $E_{K_{i(B)}}(m_{B \to A})$ corresponds to a message sent by $B$ to $A$, encrypted with $K_{i(B)}$.**

## 5. CONCLUSIONS

Motivated by the recent surge in network coding research and its applications, we discussed the security implications of stateless and state-aware network coding protocols in comparison with common routing solutions. By exploiting the intrinsic properties of network coding, most notably the algebraic dependency between packets and the possibility of information dissemination along multiple paths, it is possible to find natural lightweight solutions for distributing keys and ensuring confidentiality against eavesdroppers. The most serious security challenges posed by network coding thus seem to come from various types of Byzantine attacks, which target either the control traffic or the network code itself and demand more sophisticated security solutions than those currently in place.

Open problems include how to construct and maintain network topologies that are particularly robust against Byzantine attacks on network coding (as pursued e.g. in [D. Wang and Kschischang 2007] via limited broadcast), how to combine network coding and cryptographic primitives in the most effective way, how to exploit the properties of network coding for anonymous communication (e.g.along the lines of the *information slicing* approach proposed in [Katti et al. 2007]), and how to leverage network coding for secure multi-party computation.

As network coding protocols continue to flourish and thrive, addressing the aforementioned concerns is likely to yield a considerable number of research opportunities in network coding security.

## Acknowledgements

The authors gratefully acknowledge insightful discussions with Muriel Médard from Massachusetts Institute of Technology, USA and Ralf Koetter from Technischen Universitaet Muenchen, Germany.

Part of this work was carried out with assistance of financial support from the European Community under grant FP7-INFSO-ICT-215252 (N-Crave Project). This work was partly supported by the Fundação para a Ciência e Tecnologia (Portuguese Foundation for Science and Technology) under grants SFRH/BD/24718/2005, SFRH/BD/28056/2006 and SFRH/BD/28946/2006.